\documentstyle[aps,prd]{revtex}
\begin{document}
\title{Evolving Lorentzian Wormholes}
\author{Sayan Kar\cite{one-email}}
\address{Institute Of Physics\\ Sachivalaya Marg, Bhubaneswar 751005, India}
\author{Deshdeep Sahdev\cite{two-email}}
\address{Department of Physics \\ Indian Institute of Technology, Kanpur
208016, India}
\date{June 1995}
\twocolumn[
\maketitle
\widetext
\parshape=1 0.75in 5.5in

\begin{abstract}
Evolving Lorentzian wormholes with the required matter
satisfying the Energy conditions are discussed. Several
different scale factors are used and the corresponding
consequences derived. The effect of extra, decaying(in time)
compact dimensions present in the wormhole metric is also
explored and certain interesting conclusions are derived for the
cases of exponential and Kaluza--Klein inflation.

\end{abstract}
\vskip 0.125 in
\parshape=1 0.75in 5.5in
PACS number(s): 04.20.Gz, 04.20.Jb
\pacs{} ]

\narrowtext

\section{\bf INTRODUCTION}

The late fifties saw the emergence of the wormhole in the
seminal papers of Misner and Wheeler{\cite{mi:annphy57}} and
Wheeler {\cite{wh:annphy57,wh:acad62}}. Electric charge was
claimed to be a manifestation of the topology of a space which
essentially looked like a sheet with a handle. This was given
the name 'wormhole'. In spite of the elegance and simplicity of
the idea, interest in the Wheeler wormhole declined over years
primarily because of the rather ambitious nature of the
programme which unfortunately had little connection with reality
or support from experiment.

The wormhole remained dormant over the years with some isolated
pieces of interesting work such as the one by Ellis
{\cite{ellis:jmp73}} appearing once in a while.

The year 1988 saw a resurgence of interest in 'wormholery'--
once again in terms of a bunch of new, exotic ideas which seems
to have remained alive for quite some time now. Two separate
directions emerged -- one concerning Euclidean signature metrics
and the other related to the Lorentzian ones.

Euclidean wormholes arose in the context of Euclidean quantum
gravity {\cite{hwk:prd88,gidd:npb88}}. The focus here was to
construct a viable model for topology change where the
transition amplitude could be evaluated at least in the
saddle--point approximation. Giddings and Strominger
{\cite{gidd:npb88}} accomplished this task by deriving an exact
wormhole solution of the Euclidean Einstein equations with
matter in the form of an antisymmetric tensor field of rank
three.  They also showed that the transition amplitude for a
topology changing process from $R^{3}$ to $R^{3} \oplus S^{3}$
is significant only when the size of the wormhole throat is of
the order of Planck length.

On the other hand, interest in Lorentzian wormholes was
stimulated by the important paper of Morris and Thorne
{\cite{mt:ajp88}} where static, spherically symmetric Lorentzian
wormholes were first defined and analysed in great detail with
emphasis on their geometry and matter content. Soon after
Morris, Thorne and Yurtsever (MTY) {\cite{mty:prl88}} (and
almost simultaneously Novikov{\cite{idn:spjetp89}} ) constructed
a model time machine using static, spherically symmetric
traversable wormholes. Since the publication of the MTY paper
the existence/nonexistence of a time--machine ({\em
theoretically!} ) has been an issue of major controversy.
Questions such as the classical mechanics of billiard
balls{\cite{ekt:prd91}} and the Cauchy problem for the scalar
wave equation in the presence of closed timelike curves
{\cite{fmnekty:prd90,fm:prl91}} have been addressed in detail.
More interestingly, Kim and Thorne{\cite{kimth:prd91}} and
Hawking {\cite{hw:prd92}} have discussed at length whether
vacuum polarization effects can prohibit the occurence of closed
timelike curves.

However, one major problem with the existence of traversable
wormholes and time machines is that the matter required to
support such a geometry essentially violates the Energy
Conditions of General Relativity(GR)
{\cite{mty:prl88,mt:ajp88}}.  One can restrict such violation to
an infinitesimally small thin shell if one adopts Visser's
approach towards wormhole construction {\cite{viss:npb89}. In
fact, Visser{\cite{viss:prd89} has also shown that for non
spherically symmetric, static wormholes one can have many null
geodesics along which the Averaged Null Energy Condition is
satisfied.

On the other hand, as shown recently by one of us here, if one
shifts attention to nonstatic geometries one finds that WEC
violation can be avoided for arbitrarily small or large
intervals of time{\cite{sk:prd94}}.  In this paper, we shall
demonstrate in greater detail, through various arguments and
examples that within Classical General Relativity there exist
Lorentzian wormholes which are nonstatic and which do not
require WEC violating matter to support them.  These
wormholes,as will turn out, exist for a finite (but arbitrarily
small or large) time interval and represent evolving geometries.
During its evolution the shape of the wormhole changes in the
embedding space--the throat radius expands or contracts and the
rate of change of the embedding function increases or decreases.
One can draw an analogy between these geometries and that of the
usual FRW universe $(k = 1)$.  The spacelike sections of the
former are topologically $R \otimes S^{2}$ while those of the
latter are $S^{3}$. In the spirit of this, one can therefore
think of these spacetimes as constituting `wormhole universes'.
Other papers which deal with evolving wormholes are due to
Hochberg and Kephart {\cite{hockep:prl93}} and Roman
{\cite{rom:prd93}}. While the former discusses a possible
resolution of the horizon problem using a network of dynamic
(evolving) wormholes possibly present in the early universe, the
latter considers an evolving geometry with an inflationary scale
factor. Very recently, Wang and Letelier {\cite{wl:grqc}} have
constructed a class of evolving wormholes which they claim can
be built out of matter satisfying both the Weak and Dominant
Energy conditions. However, the approach towards constructing
evolving wormholes that we discuss in this paper are somewhat
different from theirs. It should also be mentioned that Visser
in {\cite{viss:npb89}} has discussed the stability of his
wormholes by making the throat dynamic (i.e time dependent).

We state explicitly that these evolving wormholes though
supported with normal matter do not necessarily violate the
theorem of Topological Censorship due to Friedman,Schliech and
Witt {\cite{tcens:prl93}}.  Topological Censorship essentially
states that asymptotically flat traversable wormhole spacetimes
cannot exist if matter satisfies the Averaged Null Energy
Condition (ANEC). More precisely, the theorem proves that there
must be at least one null geodesic along which the ANEC is
violated.  In many of the spacetimes to be discussed here
asymptotic flatness is not assumed. However, spacelike sections
when embedded in $R^{3}$ resemble two asymptotically flat
regions connected by a bridge. We should remember that the
essential features of a wormhole geometry are largely encoded in
the spacelike section and in the condition for nonexistence of
horizons ($g_{00} \neq 0$). Moreover, our geometries do violate
WEC in some interval of time {\em but not always}). In fact by
suitably adjusting parameters one can make the timespan over
which WEC is satisfied as large as one wants.

The paper is organised as follows. The next section deals with
the proof of the fact that evolving wormholes can satisfy the
WEC. In Section III we discuss some specific examples in $2+1$ and
$3+1$ dimensions. Section IV illustrates a specific model of a
wormhole in an FRW universe. Extra compact decaying dimensions
present in the wormhole metric are dealt with in Section V
where two specific models are presented. Finally, in Section VI we
conclude with some remarks on future directions.

\section {\bf EVOLVING WORMHOLES AND ENERGY CONDITIONS}

We begin our analysis with the following ansatzen for the metric
and the energy-momentum tensor.
\begin{equation}
ds^{2} = \Omega ^{2}(t) [- dt^{2} + {{dr^{2}}\over {1- b(r)/r}}
+ r^{2} d\Omega ^{2}_{D-2} ]
\end{equation}
\begin{equation}
T_{oo} = \rho (r,t),\quad T_{11} = \tau(r,t),\quad T_{jj} =
p(r,t)
\end{equation}

where $j$ runs from $2$ to $D-1$ and $\Omega ^{2}(t)$ is the
conformal factor, finite and positive definite throughout the
domain of $t$. One can also write the metric in (1) using
`physical time' instead of `conformal time'. This would mean
replacing $t$ by $\tau = \int \Omega (t)dt$ and therefore
$\Omega (t)$ by $R(\tau)$ where the latter is the functional
form of the metric in the $\tau$ coordinate.  However, at the
moment we use `conformal time'.  Translating all the results for
$t$ into those for $\tau$ is simple. In the fourth and fifth
sections of the paper (where we shall explicitly deal with
realistic models) we shall use ${\tau}$.  $\rho (r,t),
\tau(r,t)$ and $p(r,t)$ are  the
components of the energy momentum tensor in the frame given by
the one-form basis
\begin{eqnarray}
e^{o} = \Omega (t)dt,\quad e^{1} = {\Omega (t) dr \over
{\sqrt{1-b(r)/r}}}, {\nonumber} \\ \quad e^{j} = \Omega (t) r
\sin {\theta_{1}}\sin {\theta_{2}}
{...........} \sin {\theta_{j-2}} d{\theta_{j-1}}
\end{eqnarray}

$ d\Omega ^{2}_{D-2}$ is the line element on the (D-2)-sphere.
$b(r)$ is the usual `shape function' as defined by Morris and
Thorne {\cite{mt:ajp88}}.  It will be assumed to satisfy all the
conditions required for a spacetime to be a Lorentzian wormhole;
i.e., ${b(r)\over r} \le 1 ; {b(r)\over r} \rightarrow 0$ as $r
\rightarrow  \infty $ ; at $r = b_{o}, b(r) = b_{o} and r \geq
b_{o}$. The Einstein field equations with the ansatz (1) and (2)
turn out to be (for $8\pi G = c^{2} = 1)$.
\par
\begin{eqnarray}
\rho (r,t) = {(D-2)\over {\Omega ^{2}}}\left [{\frac{(D-1)}{2}}
{\left ({{\dot
{\Omega}} \over {\Omega}}\right )^{2}} \right . {\nonumber} \\
\left .+{b ^{\prime}r
+(D-4)b } \over {2r^{3} }\right ]
\end{eqnarray}
\begin{eqnarray}
\tau (r,t) = {(D-2)\over {\Omega ^{2}}}\left [
-{{\ddot {\Omega}} \over {\Omega}} - {\frac{(D-5)}{2}} {\left
({{\dot {\Omega}} \over {\Omega}}\right )^{2}}\right .
{\nonumber}
\\ \left . - (D-3){b \over {2r^{3}}}\right ]
\end{eqnarray}
\begin{eqnarray}
p(r,t) = {(D-2)\over {\Omega ^{2}}}\left [-{{\ddot {\Omega}}
\over {\Omega}} -
{\frac{(D-5)}{2}\left ({{\dot {\Omega}} \over {\Omega}}\right
)^{2}}
\right ] {\nonumber}\\ +
\left [{{-(D-3)(D-5)b+(3-D){b^{\prime}} r}\over {2r^{3}\Omega^{2}}} \right ]
\end{eqnarray}

The dots denote derivatives with respect to $t$ and the primes
derivatives with respect to $r$. The WEC $(T_{\mu \nu } u^{\mu }
u^{\nu } \ge 0 \forall $ nonspacelike $u^{\mu })$ reduces to the
following inequalities for the case of a diagonal
energy-momentum tensor

\begin{equation}
\rho  \ge  0,\quad \rho  + \tau \ge  0 ,\quad  \rho  + p \ge  0
\quad \forall (r,t)
\end{equation}

{}From Eqns. (4), (5), (6) one can write down three inequalities
which have to be satisfied if the WEC is not to be violated.
These are

\begin{equation}
{(D-2)\over {\Omega ^{2}}}\left [{\frac{(D-1)}{2}}{{\left
({{\dot {\Omega}} \over {\Omega}}\right )^{2}}}+{{b ^{\prime}r
+(D-4)b } \over {2r^{3}}}\right ] \ge 0
\end{equation}
\begin{equation}
{(D-2)\over {\Omega ^{2}}}\left [-{{\ddot {\Omega}} \over
{\Omega}} + 2{\left ({{\dot {\Omega}} \over {\Omega}}\right
)^{2}} - {{b-{b^{\prime}} r}\over {2r^{3}}} \right ] \ge 0
\end{equation}
\begin{equation}
{(D-2)\over {\Omega ^{2}}}\left [-{{\ddot {\Omega}} \over
{\Omega}} + 2{\left ({{\dot {\Omega}} \over {\Omega}}\right
)^{2}}\right ] +{{(2D-7) b+{b^{\prime}} r}
\over {2r^{3}\Omega^{2}}} \ge 0
\end{equation}

Several important facts should be noted here in comparison with
the case of a static geometry.  Eq. (8) is satisfied if
$b^\prime \ge 0$ and $D\ge 4$ irrespective of the geometry being
time independent/dependent. However, if it is time dependent,
then one can satisfy Eq (8) even for the case when $b^{\prime}
\le  0$.  For example, for $D=4$
one obtains the inequality

\begin{equation}
{{\vert {b^{\prime}} \vert}\over r^{2}} \le 3{\left({\dot \Omega
\over \Omega}
\right)}^{2}
\end{equation}

For every $t=constant$ slice Eq. (11) has to hold, which means
\begin{equation}
{{\vert {b^{\prime}} \vert}\over r^{2}} \le min \left [
3{\left({\dot \Omega
\over \Omega}\right)}
^{2}\right ]
\end{equation}
where $min$ denotes the minimum value of the function in the
given time interval.

For a static geometry Eq. (9) can never be satisfied, as shown
by Morris and Thorne {\cite{mt:ajp88}}. But, for a nonstatic
geometry with $b^\prime \ge 0$ one {\em can} satisfy Eq.(9). We
require
\begin{equation}
\left [{2{\left ({{\dot {\Omega}} \over {\Omega}}\right )^{2}}-
{{\ddot {\Omega}} \over {\Omega}}}\right ] \ge {{b-{b^{\prime}}
r}\over {2r^{3}}}
\end{equation}

Interestingly, Eq. (13) is the same in all dimensions.  Stated
explicitly Eq (13) implies that the value of $(b-b^\prime
r)/r^{3}$ for all $r$ must be $_{.}$less$_{.}$\d{t}han or equal
to the minimum value of the function $\left [{2{\left ({{\dot
{\Omega}} \over {\Omega}}\right )^{2}}- {{\ddot {\Omega}} \over
{\Omega}}}\right ] $ , in the corresponding domain of $t$.
However, we need

\begin{equation}
F(t) =\left [{2{\left ({{\dot {\Omega}} \over {\Omega}}\right
)^{2}}- {{\ddot {\Omega}} \over {\Omega}}}\right ] > 0
\end{equation}

Eq (14) can be written in a more precise form by introducing a
function $\chi (t) = \Omega /\dot{\Omega }$ . We have

\begin{equation}
{d\chi \over dt} > (-1)
\end{equation}

With $b^\prime \ge 0$, $D\ge 4$ and Eq (13) holding one clearly
sees that Eq.(9) is satisfied.  Therefore from this very simple
analysis it is clear that nonstatic spherically symmetric
Lorentzian wormhole geometries can exist with the required
matter not violating the WEC. However, the fact that $\Omega
(t)$ be finite everywhere and must satisfy the condition Eq.
(13) implies that these wormholes exist for finite intervals of
time (arbitrarily small or large ). For finite and bounded
$\Omega$ which is everywhere nonzero this can be proved as
follows.The finiteness and boundedness of $\Omega (t)$ implies
that it can have only a specific class of functional forms.
These include (i) functions which have no extremum and
asymptotically approach the constant limiting values (e.g.
$A+B\tanh {\omega t} $) (ii) functions which have one extremum
and asymptotically approach constant limiting values (e.g.
$A+Be^{-\omega t^{2}}$) (iii) oscillatory functions which may or
may not approach the limiting values (e.g $\sin {\omega t} + a
$). In all the three cases where the functions asymptotically
approach limiting values $F(t)$ tends to zero at $t\rightarrow
\pm \infty $ and the WEC is violated.
For a purely oscillatory function there exists more than one
extremum and at the minimum $F(t)$ is clearly negative. Thus WEC
violations can be avoided only for finite intervals of time if
one chooses a finite and bounded $\Omega (t)$.

One can also derive the fact that the WEC will be violated for
some interval of time by the following argument (this was
communicated to us by Matt Visser {\cite{viss:prv}}).

Consider the condition $F(t) > 0$. This can be written
equivalently as

\begin{equation}
F(t) = \Omega \frac{d^{2}}{dt^{2}}({\Omega}^{-1}) > 0
\end{equation}

Now assuming a well behaved $\Omega$ we can evaluate the
integral of $F(t){\Omega}^{-2}$. This gives

\begin{equation}
{\int}_{-\infty}^{+\infty}F(t)\Omega^{-2} dt =
{-\int}_{-\infty}^{+\infty}
\left[\frac{d}{dt}(\Omega^{-1})\right] dt < 0
\end{equation}

on integration by parts.

Since the integral is negative, the integrand must be negative
somewhere.  Therefore $F(t)$ cannot be positive everywhere if
$\Omega$ is well behaved everywhere. Thus the WEC must be
violated somewhere atleast.

In $2+1$ dimensions the WEC conditions turn out to be

\begin{equation}
\frac{1}{\Omega^{2}}\left [ {\left (\frac{\dot \Omega}{\Omega}\right)}^{2}
+ \frac{b^{\prime}r - b}{2r^{3}} \right ] \ge 0
\end{equation}
\begin{equation}
\frac{1}{\Omega^{2}}\left [ 2{\left (\frac{\dot \Omega}{\Omega}\right)}^{2}
-\frac{\ddot \Omega}{\Omega}+ \frac{b^{\prime}r - b}{2r^{3}}
\right ] \ge 0
\end{equation}

In order to satisfy both these inequalities we need the
following to hold:

\begin{equation}
\frac{d}{dt} \left ( \frac{\dot \Omega}{\Omega} \right ) \le min [\rho(r,t)]
\end{equation}
\begin{equation}
min {\left (\frac{\dot \Omega}{\Omega}\right)}^{2} \ge \frac{b -
b^{\prime}r }{2r^{3}}
\end{equation}

Some special exact solutions with simple equations of state will
be discussed in next section.

We now move on to examine how the geodesic focussing arguments
for WEC violation by static wormholes get modified once the
latter are allowed to evolve. We recall that the expansion
$\theta$ of a congruence of null rays satisfies the Raychaudhuri
equation (the rotation $\omega_{\mu\nu}$ has been set to zero
{\cite{akr:pr55,wald:gr85}}):

\begin{equation}
{\frac{d\theta}{d\lambda}}+\frac{1}{2}{\theta}^{2} =
-R_{\mu\nu}{\xi}_{\mu} {\xi}_{\nu} -2{\sigma}^{2}
\end{equation}

Here ${\sigma}_{\mu\nu}$ is the shear tensor of the geodesic
bundle (which is zero for our case here). From Einstein's
equation we know that $R_{\mu\nu}
{{\xi}^{\mu}}{{\xi}^{\nu}}=T_{\mu\nu}{{\xi}^{\mu}}{{\xi}^{\nu}}
$ for all null ${\xi}^{\mu}$. So if
$T_{\mu\nu}{\xi}^{\mu}{\xi}^{\nu} \ge 0$, ${\theta}^{-1}\ge
{\theta}_{0}^{-1}+\frac{\lambda}{2}$ by (9). Thus if $\theta$ is
negative anywhere it has to go to $-\infty$ at a finite value of
the affine parameter, $\lambda$, i.e. the bundle must
necessarily come to a focus.  In the case of a static wormhole
,the expression for $\theta$ is {\cite{soma:prd92}}

\begin{equation}
\theta = 2\beta \frac{r^{\prime}}{r}
\end{equation}

where $\beta$ is a positive quantity and
$r^{\prime}=\frac{dr}{dl}$ and $l$ is defined by

\begin{equation}
l(r)=\pm \int_{b_{0}}^{r}{\frac{dr}{\sqrt{1-\frac{b}{r}}}}
\end{equation}

Thus if $r^{\prime}$ is negative anywhere, and it is so for
$l<0$, then so is $\theta$. But $\theta \rightarrow -\infty$
only if $r \rightarrow 0$, since $r^{\prime}$ is always finite.
Hence either the wormhole has a vanishing throat radius which is
tantamount to its not being a wormhole at all,or the WEC is
violated.

For the evolving case $\theta$ is given by \begin{equation}
\theta = \frac{2\beta}{R(\tau)}\left ({\tilde {R(\tau)}} +
\frac{r^{\prime}}{r}\right ) \end{equation}

Here we have used real time $\tau$ and the $\tilde {}$ denotes
differentiation with respect to $\tau$.  Thus as long as
${\tilde {R}} >{\vert \frac{r^{\prime}}{r}\vert}$ i.e.  the
wormhole is opening out fast enough , $\theta$ is never negative
and the fact that the bundle does not focus no longer implies
WEC violation.  We emphasize that $\theta >0$ while being a
necessary condition for WEC conservation is not sufficient.  To
see this let us assume $b(r)=\frac{b_{0}^{2}}{r}$ and
$R(\tau)=\alpha \tau$, the scale factor for a Milne universe.
For these choices of $b(r)$ and $R(\tau)$ the condition $\rho
\ge 0$
leads to $b_{0}^{2}{\alpha}^{2} \ge \frac{1}{3}$. However
(18),(19) leads to $b_{0}^{2}{\alpha}^{2}\ge \frac{1}{2}$ which
is more stringent in the sense that if it be satisfied then the
$\rho \ge 0$ inequality is trivially true.  As for $\theta $ ,
$\frac{r^{\prime}}{r}$ passes through a minimum at
$l=\frac{-1}{2b_{0}}$ (note that $r^{2}(l)=b_{0}^{2}+l^{2}$)
while $\tilde R =
\omega $. Clearly if $\alpha >\frac{1}{{\sqrt{2}}b_{0}}$, $\theta >0$
, as we would expect for WEC preservation. On the other hand ,
if we choose $R(\tau)=e^{{\alpha}\tau}$, the inflationary scale
factor, and require that ${\alpha} > {\frac{1}{2b_{0}}}$, then
$\theta >0$ but (19) is no longer satisfied as
$\frac{d}{d\tau}\frac{\tilde R}{R} =0 $ identically.

In other words, the focussing argument which is- `$\theta <0
\Rightarrow $
WEC Violation for wormholes' is useful for picking out WEC
violation ,but not WEC conservation. For this it is necessary to
examine $T_{\mu\nu}{{\xi}^{\mu}}{{\xi}^{\nu}}$ explicitly as
done above.

Before we construct explicit examples it is useful to discuss
briefly the embedding in $R^{3}$ of a $\theta = \pi /2$, $t$ =
$t_{0}$ slice, where $t_{0}$ lies in the interval in which the
wormhole exists.  Since our geometry is nonstatic each such
slice will be different - more precisely the value of the
function $\Omega (t)$ at $t = t_{0}$ will dictate the shape and
features of this slice, which will thus change as we alter
$t_{0}$. The metric on such a slice takes the form
\begin{equation}
{\bar {ds}}^{2} = {\Omega}^{2}(t_{0})\left [{dr^{2}\over
{1-{b(r)\over r}}} + r^{2}{d\phi}^{2}\right ]
\end{equation}
Define
\begin{equation}
\tilde r = \Omega (t_{0}) r
\end{equation}
Thus the metric on the slice takes the form
\begin{equation}
{\bar {ds}}^{2} = {{d{\tilde r}^{2}} \over {1-{{{a(\tilde
r})\Omega(t_{0})}\over {\tilde r}}}} +{\tilde r}^{2}{d\phi}^{2}
\end{equation}
where $a(\tilde r)$ is the functional form of $b(r)$ in the
$\tilde r $ coordinate.  The minimum value of $\tilde r$ which
determines the throat radius is evaluated from
\begin{equation}
a(b_{0})\Omega (t_{0}) = \tilde b_{0}
\end{equation}
This clearly shows the dependence on $\Omega (t) $.Using the
mathematics of embedding we can write the following differential
equation for the spacelike slice at $ t= t_{0}$.
\begin{equation}
{dz({\tilde r})\over d{\tilde r}} = \pm {\left [{ a({\tilde
r})\Omega (t_{0}) }\over {{\tilde r} - a( {\tilde r})\Omega
(t_{0})}\right ] }^{1\over 2}
\end{equation}
where $z({\tilde r})$ is the embedding function.  Integrating
(28) one can obtain the $z({\tilde r})$ for the slice at $t =
t_{0}$.

\section { \bf EXAMPLES}

\subsection {\bf 2+1 Dimensions}

As mentioned in the previous section, the $2+1$ dimensional case
is somewhat special. Here we present two exact evolving wormhole
solutions.
 The important point to realise is that in $2+1$ dimensions the
$\tau = p$ constraint is automatically satisfied.

We first discuss the case $\tau = p = 0$.  From the field
equations it turns out that the conformal factor $\Omega (t) =
\exp {\pm \omega t}$ is the only solution that satisfies this
constraint. This scale factor in real time is the familiar Milne
universe in $2+1$ dimensions.  The only WEC inequality we have
to check here is the one for $\rho$ (i.e. $\rho \ge 0$).
Choosing $b(r) = b_{0}$ we end up with a requirement

\begin{equation}
b_{0}^{2}{\omega}^{2} \ge \frac{1}{2}
\end{equation}

On the other hand if we choose $\tau = p =p_{0} > 0$ (where
$p_{0}$ is a constant) the solution to the differential equation
governing the evolution of the conformal factor is:

\begin{equation}
\Omega (t) = \frac{1}{\cosh t\sqrt{p_{0}}}
\end{equation}

A requirement for WEC conservation is

\begin{equation}
\frac{b_{0}}{2r^{3}} \le p_{0} \tanh^{2} {\sqrt{p_{0}} t}
\end{equation}

Thus one has to allow WEC violation atleast for a small interval
$(-t_{0}, t_{0})$ in the neighborhood of $t = 0$. The resulting
constraint on the throat radius parameter $b_{0}$ becomes

\begin{equation}
b_{0}^{2} \ge \frac{1}{2p_{0}\tanh ^{2}t_{0}\sqrt {p_{0}}}
\end{equation}

\subsection {\bf 3+1 Dimensions}

In all the examples to be discussed here we shall choose $b(r)
=b_{0}$.  The WEC imposes the following restriction on $b_{0}$:
\begin{equation}
b_{0}^{2} \ge max {\left ({1\over {F(t)}}\right )}
\end{equation}
This condition will restrict the minimum possible value of the
throat radius of the wormhole. Other choices of $b(r)$ lead to a
modified version of the above stated condition and of the
subsequent analysis.
\par
\medskip
(i) \, $\Omega (t) = {1\over {C - \omega t}} $
\medskip

This choice of $\Omega $ was first discussed by Roman
{\cite{rom:prd93}}.  It is essentially the case of an
inflationary wormhole universe.Roman's idea was to build a model
of a wormhole of large size emerging out of inflation.One can
check very easily that this choice of the scale factor leads to
$ F(t) = 0$ , which in turn implies that the WEC will be
violated for all times.More details about this scenario can be
found in the paper by Roman {\cite{rom:prd93}}.

\medskip
(ii) \, $\Omega (t) = \exp (\pm {\omega t}) $
\medskip

In a sense this case is also quite unique because it leads to an
$F(t) = 2{\omega}^{2} $, which is a constant. This scale factor
is reminiscent of the one used for the Milne universe.In real
time it represents linearly expanding spacelike slices.The WEC
is satisfied if and only if the following inequality holds:
\begin{equation}
b_{0}^{2}{\omega}^{2} \ge {1\over 2}
\end {equation}
The time interval for which the wormhole can exist ( assuming
that it can grow very large) is $ - \infty < t \le \infty $ for
the $\exp ({\omega t})$ case and $ - \infty \le t < \infty $ for
the $ \exp (-{\omega t})$ case.Note that one of the infinities
is excluded from each of the intervals because the spacetime
collapses into a singularity there.

\medskip
(iii) \, $\Omega (t) = \sin {\omega t} $
\medskip

The scale factor is the same as the one that arises in the
closed FRW cosmology which starts with a `bang' and ends with a
`crunch'.Instead of the usual $S^{3}$ spacelike sections we have
wormhole metrics on $ R\otimes S^{2} $ .The expression for $F(t)
$ turns out to be
\begin{equation}
F(t) = 2{\omega}^{2}( 2\cot ^{2}{\omega t} + 1)
\end{equation}
One can easily check that $F(t)$ has a minimum at $ \omega t =
{\pi \over 2} $ .Hence the constraint on the allowed values of $
b_{o}$ turns out to be the same as in the previous case i.e $
b_{0}^{2} {\omega}^{2} \ge {1\over 2} $ .One can also say that
the lifetime of this wormhole universe is $\pi \over {\omega}$
.The time interval for which this universe can exist
without collapsing into a singularity is $ {m\pi \over
{\omega}}< t < {(m+1)\pi \over {\omega}} $.

\medskip
(iv) \, $\Omega (t) = {(\omega t)}^{\nu} $\qquad $\nu$ integral
or fractional
\medskip

This case is important because the scale factors that arise in
the dust-filled or radiation dominated FRW cosmologies with flat
spacelike sections are obtained by considering special values of
the $\nu$ used above.  We shall deal with these special cases
later in a separate section.

For general $\nu $ the expression for $F(t)$ is given as
\begin{equation}
F(t) = {{2\nu (\nu +1)}\over {t^{2}}}
\end{equation}
Therefore as $t \rightarrow \pm \infty $ $F(t) \rightarrow 0$.
The constraint on $b_{0}$ turns out to be dependent on $t$ .
\begin{equation}
b_{0}^{2} \ge max{\left ({t^{2}\over {2\nu(\nu +1)}}\right )}
\end{equation}
Thus the evolving wormhole with this type of scale factor can
exist only for a finite interval of time ( similar to the
previously discussed cases ). The lower bound on $b_{0}$ is
decided by the maximum time $t$ upto which we wish the wormhole
to exist with the matter threading the geometry satisfying the
WEC.Beyond this time the matter will violate the WEC.  As we
shall see later this condition sets the most stringent
restriction on the size of the wormhole.

We now move on to a special class of scale factors which exhibit
`flashes' of WEC violation. This essentially means that the
matter threading the wormhole violates the Energy Conditions
only for small intervals of time and is normal at all other
times. Whether such matter is physically possible is as yet
unknown. However , the intervals during which the WEC is
violated can be chosen to be very small (this obviously leads to
a constraint on the throat radius). We shall deal with two
representative examples. Many more can constructed without much
of a problem.

(a) \, $\quad \Omega (t) = \sin {\omega t} + a \qquad a>1$

This scale factor reminds us of the `bounce' type solutions in
cosmology which were constructed in order to avoid the big--bang
singularity.  For a nonstatic wormhole with the scale factor
chosen as above the function $F(t)$ turns out to be :
\begin{equation}
F(t) = 2{\omega}^{2}\left [{{2-{\sin}^{2}{\omega t} + a\sin
{\omega t}}\over {({\sin \omega t +a})^{2}}}\right ]
\end{equation}

This function is plotted in Fig 1.

Note that $F(t)$ goes to zero in the vicinity of $\omega t =
3\pi /2$ ,stays negative for a while and then subsequently
becomes positive again.This happens once in every period , i.e
at $\omega t$ values given by $(2n+1)\pi /2$.Thus the WEC will
be violated during those intervals of time.The time domain over
which the WEC is violated is ,as mentioned before , related to
the throat radius parameter of the wormhole.One chooses this
interval in the following way.  Assume a minimum ,positive and
finite value of $F(t)$ ,say $F_{0}$.In the neighborhood of the
point where $F(t)$ is zero , this value occurs at points say
$\omega t_{0} -\delta$ and $\omega t_{0} +\delta $.If the shape
function is assumed to be constant ($b(r) = b_{0}$) then the
relation between the throat radius parameter and the minimum
value of $F(t)$ turns out to be :
\begin{equation}
b_{0}^{2} \ge {1\over F_{0}}
\end{equation}
Hence the interval during which the WEC is violated is $2\delta$.
The smaller this time interval the larger the minimum allowed
value of $b_{0}$.

\medskip
(b)\quad $\Omega (t) = {{t^{2} + a^{2}}\over {t^{2} + b^{2}}}$
\qquad $b>a>0$
\medskip

In this case also the scale factor is such that the geometry is
never singular.Asymptotically it becomes a static wormhole.The
function $F(t)$ turns out to be

\begin{equation}
F(t) = {{4(b^{2}-a^{2})(3t^{2}-a^{2})}\over
{(t^{2}+b^{2})({t^{2}+a^{2}})^{2}}}
\end{equation}

This function is shown plotted in Fig 2. At $t=\pm a/{\sqrt 3}$
$F(t) = 0$.For $-a/{\sqrt 3} < t < a/{\sqrt 3}$ $F(t)$ is
negative.One can carry out an analysis similar to the one for
(a)--the only difference being that WEC violation occurs here
only in the neighborhood of the interval mentioned above and at
$ \pm \infty $.

Further examples can be constructed by choosing other forms of
$\Omega (t) $. Two worth mentioning follow from constraints on
the matter stress energy for an evolving wormhole geometry. For
the perfect fluid with $p={\rho \over 3}$ we end up with the
scale factor of a closed FRW universe while for traceless matter
in general i.e matter obeying only $-\rho +\tau +2p = 0$ we get
a linear $\Omega (t)$ i.e $\Omega (t)=at+b $.

\section {A WORMHOLE IN A FLAT FRW UNIVERSE}

A probable realisation of an evolving wormhole could be obtained
by thinking of it as part of an asymptotically FRW universe i.e.
by imagining the asymptotically flat parts of an evolving
wormhole geometry as constituting a flat ($k=0$) FRW spacetime.
Mathematically one therefore chooses a metric which represents
an evolving wormhole with the scale factor identical to that of
either the matter or the radiation dominated FRW model. Thus
\begin{equation}
ds^{2} = -d{\tau}^{2} + {\tau}^{n}\left ( {dr^{2}\over
{1-{b(r)\over r}}} + r^{2}d{\Omega}_{2}^{2}\right )
\end{equation}

where we have changed our time coordinate from the conformal
time used in the earlier discussions to real time. The exponent
$n$ takes on the values $1\over 2$ and $2\over 3 $ for the
radiation and matter dominated cases of the flat FRW universes
respectively. The Einstein equations lead to the following
expressions for the matter density and the pressures.
\begin{equation}
\rho (r,\tau) = {{b^{\prime}}\over {({\tau})^{2n}r^{2}}} + {{3n^{2}}\over
{({c\tau})^{2}}}
\end{equation}
\begin{equation}
p_{1}({r,\tau}) = -{b\over {{({\tau})^{2n}}r^{3}}} -
{{n(3n-2)}\over {({c\tau})^{2}}}
\end{equation}
\begin{equation}
p_{2}(r,\tau) = {{b-b^{\prime}r}\over {{2r^{3}}{({\tau})^{2n}}}}
- {{n(3n-2)}\over {({c\tau})^{2}}}
\end{equation}

The WEC inequality $\rho + p_{1} \ge 0 $ reduces to the
following :

\begin{equation}
{{b^{\prime}r - b}\over {{2r^{3}}{({\tau})^{2n}}}} + {{2n}\over
{({c\tau})^{2}}} \ge 0
\end{equation}

As expected all the essential properties of the matter stress
energy of the flat FRW model follow from the expressions for
$\rho$, $p_{1}$ and $p_{2}$. As $r \rightarrow \infty $ , only
the $\tau$ dependent terms survive, and we get $\rho =
{{\rho}_{FRW}} = {4\over {3({c\tau})^{2}}}$, $p_{1}= p_{2} = 0$
for dust and ${\rho \over 3} = p_{1} = p_{2} = {1\over
{4({c\tau}) ^{2}}} $ for pure radiation.

Secondly , the pressures become increasingly disparate as we
approach the throat ($p_{2} -p_{1} = {{3b-b^{\prime}r}\over
{2{({\tau})^{2n}}r^{3}}} \ge 0 $).From the geometrical viewpoint
this is because the curvature s in the $\theta \phi $ directions
is different from that in the $r$ direction.Thus, inside the
wormhole, matter is subject to anisotropic stresses and to
remain in equlibrium it must respond by generating anisotropic
strains.

Lastly let us look at the WEC condition. For concreteness and
simplicity we choose $b(r) = b_{0}$. From the WEC inequality
$\rho + p_{1} \ge 0 $ we arrive at the condition $b_{0}R(\tau)
\ge {({c\tau})\over
{\sqrt 2n}}$.This quite simply leads to the statement that the
throat radius ${b_{0}}R(\tau)$ at time $\tau$ must exceed the
horizon size $c\tau$ (upto a constant factor given by $\sqrt
{1/2n}$ which is of O(1),being exactly equal to $1$ and ${\sqrt
3}\over 2$ for pure radiation and dust respectively).  Now for
an FRW evolution the horizon always moves faster that the scale
$R(\tau)$ and hence eventually overtakes $b_{0}R(\tau)$, causing
WEC violation to begin at the throat and slowly spread
outward.Furthermore, since wormholes presumably arise through
quantum gravitational processes, they can reasonably be assumed
to form in the Planckian era,with a radius ,which is typically
of Planck size at Planck time. If their subsequent evolution is
FRW--like ,WEC violation occurs within a few Planck times.To
avoid this ,the wormhole must inflate to a size much larger than
the horizon.This is precisely what happens in the early universe
(inflationary epoch).However, as we have shown earlier an
inflationary scale factor leads to WEC violation at all times.

\section{THE CASE OF COMPACT EXTRA DIMENSIONS}

\subsection{\bf Exponential Inflation}

It is well known from the work of Roman {\cite{rom:prd93}} and
from the analysis presented in the previous sections that with
an inflationary scale factor one cannot avoid the violation of
the WEC even for a finite interval of time . This is somewhat
depressing, because if one believes in the existence of
wormholes, then one possible way in which they can appear on
macroscopic scales is by growing very large during the
inflationary epoch. We now demonstrate through an example that
with compact extra dimensions ($S^{2}$) an inflationary wormhole
can exist for a finite interval of time with matter satisfying
the WEC.

The assumption for the metric is :
\begin{eqnarray}
ds^{2}=-dt^{2} + a_{1}^{2}(t)\left ( \frac{dr^{2}}{1-b/r}
+r^{2}d{\Omega}^{2}_ {2}
\right ){\nonumber} \\+ a_{2}^{2}(t) \left (d{\chi}^{2} + \sin^{2} \chi
d{\xi}^{2} \right )
\end{eqnarray}

where $a_{1}(t)$ and $a_{2}(t)$ are the scale factors for the
wormhole and the compact extra dimensions respectively. We
assume

\begin{equation}
a_{1}(t) = e^{\omega t}, \quad a_{2}(t) = (t_0-t)^{1/2} , \quad
b(r)=b_{0}
\end{equation}

Instead of writing down the Einstein equation in all its glory,
we straightaway move on to the WEC inequalities.
$\rho$,$\tau$,$p_{1}$,$p_{2}$, $p_{3}$ and $p_{4}$ are the six
non--zero diagonal components of the energy momentum tensor. We
have essentially four inequalities as $p_{1}=p_{2}$ and
$p_{3}=p_{4}$ by symmetry.

\begin{equation}
\rho \ge 0 \Rightarrow  3{\omega}^{2} + \frac{1-3\omega}{t_{0}-t} +
\frac{1}{4({t_{0}-t})^{2}} \ge 0
\end{equation}
\begin{equation}
\rho +\tau \ge 0 \Rightarrow - \frac{b_{0}e^{-2\omega t}}{r^{3}}
+\frac{2}{(t_{0}-t)}\left [-\omega + \frac{1}{2(t_{0}-t)} \right
] \ge 0
\end{equation}
\begin{equation}
\rho +p_{1} \ge 0 \Rightarrow  \frac{b_{0}e^{-2\omega t}}{2r^{3}}
+\frac{2}{(t_{0}-t)}\left [-\omega + \frac{1}{2(t_{0}-t)} \right
] \ge 0
\end{equation}
\begin{equation}
\rho +p_{3} \ge 0 \Rightarrow -3{\omega}^{2} + \frac{2-3\omega}{t_{0}-t} +
\frac{1}{2({t_{0}-t})^{2}} \ge 0
\end{equation}

The quantity reminiscent of the $F(t)$ is the second term in the
second and third inequalities. If this term is positive (which
is possible if $t>t_{0} -\frac{1}{2\omega}$ ) then the third
inequality is satisfied and the second one yields a bound on the
allowed domain of $b_{0}$. The first inequality is satisfied for
all $\omega \le \frac{1}{3}$ while the fourth one gives another
lower bound on $t$ :
\begin{equation}
t \ge t_{0} - \frac{{\sqrt {(2-3\omega)^{2} +24{\omega}^{2}}} +
(2-3\omega)}{12 {\omega}^{2}}
\end{equation}

Of the two lower bounds on $t$ one has to choose the more
stringent one.  For instance if $\omega =\frac{1}{3}$ then the
first bound gives $t \ge t_{0} -1.5$ and the second one implies
$t \ge t_{0} - 2.2$ .Thus , if we assume $t \ge t_{0} -1.5 $ the
other inequality is automatically satisfied.  The wormhole can
exist for the interval $t_{0}-1.5 < t < t_{0}$ with matter
satisfying the WEC. The bound on the throat radius is:
\begin{equation}
b_{0}^{2} \ge max\left [ \frac{e^{-2\omega
t}(t_{0}-t)^{2}}{{\omega}({t-t_{0} }+1/2\omega}) \right ]
\end{equation}

Thus we have indicated through the above example that an
inflationary wormhole with compact extra dimensions can exist
for a finite interval of time with matter satisfying the WEC.

\subsection{\bf Kaluza Klein Inflation}

We now move on to a more realistic case of a Kaluza--Klein
cosmology.  Here, at early times, we have a Kaluza--Klein
universe which is an exact solution of the field equations in
higher dimensional GR .  The wormhole, as we shall see in this
case, can comfortably inflate without doing any violence to the
WEC. To see how this works out, we assume the topology of the
universe to be $R\otimes W \otimes S^{D}$ , where $R\otimes W$
refers to the wormhole spacetime and the $S^{D}$ corresponds to
an extra dimensional 2--sphere. The metric describing such an
universe has the form:

\begin{eqnarray}
ds^{2} = -dt^{2} + a_{1}^{2}(t)\left [
\frac{dr^{2}}{1-{\frac{b(r)}{r}}}
+r^{2}d{\Omega}_{2}^{2}\right ] {\nonumber} \\ + a_{2}^{2}(t)
d{\Omega}^{2} _{D}
\end{eqnarray}

where $a_{1}(t)$ and $a_{2}(t)$ are the scale factors associated
with the wormhole and the compact D--sphere respectively.

{}From the Einstein Field Equations we arrive at the following
four energy condition inequalities:

\begin{equation}
3{\left ( \frac{\dot a_{1}}{a_{1}}\right )}^{2} + 3D{\frac{{\dot
a_{1}} {\dot a_{2}}}{a_{1}a_{2}}} + {\frac{D(D-1)}{2}}\left
[{\left ( \frac{\dot a_{2}} {a_{2}}\right )}^{2} + {\frac
{1}{a_{2}^{2}}}\right ] + {\frac{b^{\prime}}{r^{2}a_{1}^{2}}}
\ge 0
\end{equation}

\begin{equation}
-2{\frac{\ddot a_{1}}{a_{1}}} + 2{\left ( \frac{\dot
a_{1}}{a_{1}}\right )} ^{2} + D{\frac{{\dot a_{1}} {\dot
a_{2}}}{a_{1}a_{2}}} -D{\frac{\ddot a_{2}} {a_{2}}} +
{\frac{b^{\prime}r - b }{r^{3}a_{1}^{2}}} \ge 0
\end{equation}

\begin{equation}
-2{\frac{\ddot a_{1}}{a_{1}}} + 2{\left ( \frac{\dot
a_{1}}{a_{1}}\right )} ^{2} + D{\frac{{\dot a_{1}} {\dot
a_{2}}}{a_{1}a_{2}}} -D{\frac{\ddot a_{2}} {a_{2}}} +
{\frac{b^{\prime}r +b }{2r^{3}a_{1}^{2}}} \ge 0
\end{equation}

\begin{equation}
-(D-1){\frac{\ddot a_{2}}{a_{2}}} - 3{\frac{\ddot a_{1}}{a_{1}}}
+ 3{\frac{{\dot a_{1}} {\dot a_{2}}}{a_{1}a_{2}}} -(D-1)\left
[{\left ( \frac{\dot a_{2}}{a_{2}}\right )}^{2} + {\frac
{1}{a_{2}^{2}}}\right ] \ge 0
\end{equation}

We can explicitly check that these are satisfied, e.g., for
$b(r)=b_{0}$ and for perfect fluid matter obeying $p=(\gamma -
1)\rho $ asymptotically.

The exact solution for $\gamma =2 $ {\cite{ds:prd85}viz.

\begin{equation}
dt = {\frac{{a_{1}(\tau) d\tau}}{\sqrt {D-1}}}, \quad
a_{1}^{d}(\tau) = \kappa_{1}{\tan}^{a} {\frac{\tau}{2}}, \quad
a_{2}^{D-1}(\tau) = \kappa_{2}{\frac{\sin
{\tau}}{{\tan}^{a}{\frac{\tau}{2}}}}
\end{equation}

with $0\le a^{2} \le {\frac{dD}{d+D-1}}$ and $0\le\tau \le \pi$
(d and D being the number of normal and extra dimensions
respectively and the $\kappa_{i}$ are arbitrary constants) is
generic in its behaviour as far as higher dimensional
cosmologies are concerned.

It is easily verified that each of the $b(r)$ independent terms
in the WEC inequalities stated above reduce to
$[1-{\frac{2}{3}}a^{2}]/{\sin^{2}{\tau}}$ for $d=4$ and $D=2$.
The first, third and the fourth inequalities are then
automatically satisfied whereas the second one constrains the
allowed values of $b_{0}$.

It follows then that wormholes can inflate and subsequently
evolve through the FRW era all the way to the present, without
violating the WEC . Such wormholes must however be as large as
the visible universe today.  This conclusion, though
observationally dissapointing is, by itself not fatal because we
are dealing with an asymptotically flat FRW universe, which
being infinite can accomodate an infinite number of visible
universes.

It is nevertheless, of interest in this context to enquire if an
oscillatory $R(t)$ (which can, in principle keep the throat from
growing indefinitely) is consistent with WEC preservation. The
answer is clearly no, if we require $R(t)$ to go through a
minimum {\em smoothly}. Indeed, at such a minimum $\dot R = 0$
and $\ddot R > 0$ which implies that $-\frac{d}{dt}
\left (\frac{\dot R}{R}\right ) < 0$ and that, consequently, the L.H.S
of the second WEC inequality cannot be positive for all $r$ and
$t$. On the other hand, if we let $R(t)$ have a cusp at the
minimum ----$R(t) = A + \sin{\omega t}$ is a concrete example
--- the WEC can be satisfied for all values of $t$. The
discontinuities in $\dot R (t)$ inherent in such choices of
$R(t)$ are of course unphysical but their existence indicates
that WEC violation {\em can} be restricted to arbitrarily small
intervals of time (and, equally important perhaps, to length
scales wherequantum gravity is important). This is, infact the
temporal analogue of the spatial confinement of exotic matter to
an infinite simally thin shell at the joint between two
asymptotically flat spaces in wormholes obtained by suturing two
such spacetimes together.  The important difference is that in
this case the energy--time uncertainty relation can be invoked
to allow arbitrarily energy violations for the delta function
time intervals in question. The possibilities multiply if we
introduce extra dimensions. The scale factor, $a_{1}(t)$ , can
now infact pass smoothly through a minimum since $\dot a_{1} (t)
= 0$ reduces the time--dependent factor in the second WEC
inequality to $-D\frac{\ddot a_{2}}{a_{2}} - 2\frac{\ddot
a_{1}}{a_{1}}$, and the $-D\frac{\ddot a_{2}}{a_{2}}$ term can,
for appropriate choice, of $a_{2}$, overcome the negative
contribution from $-2\frac{\dot a_{1}}{a_{1}}$.  However, if a
wormhole with an oscillating throat is simultaneously part of an
asymptotically flat FRW universe, then its scale factor,
$R(r,t)$ , must necessarily be $r$ as well as $t$ dependent.
This $r$ dependence is also necessary to keep the slowing down
in the expansion of the asymptotic region from causing the WEC
violation which it inevitably would, if the entire space were
governed by a single scale factor. For a scenario, involving
extra dimensions the oscillations in $a_{1}(r,t)$ at the throat
could plausibly be coupled to those in $a_{2}(r,t)$; indeed they
could well be the result of such a coupling. A complete theory
of astrophysical wormholes based on $r$ and $t$ dependent scale
factors, which exploits these possibilities will be reported
elsewhere.

\section {CONCLUDING REMARKS}

We have shown in this paper that wormholes with normal matter
are a realistic possibility even in the domain of classical GR.
The existence of these geometries for a finite interval of time
with matter satisfying the WEC seems a little disturbing
although it is definitely better than the situation for static
geometries. Consequences of the presence of such an evolving
wormhole in the flat FRW model have been outlined in brief.  The
role of extra compact decaying dimensions have also been dealt
with in the context of two simple models --one involving an
exponential inflation and the other a Kaluza--Klein type
inflation..

It is quite possible that the matter required for spacetimes
which exhibit 'flashes' of WEC violation might actually satisfy
the Averaged Null or Averaged Weak Energy Conditions. The reason
to believe in this is related to the fact that the Averaged
Energy Conditions are global (one integrates over the quantity
$T_{\mu\nu}{\xi}^{\mu}{\xi}^{\nu}$ along a certain timelike/null
geodesic). However we have not carried out this calculation
although we hope to do it in due course.

We have also not dealt with the question of human traversability
in the context of these evolving geometries. Such an analysis,
is however, not too diffi cult. One has to replace the static
observer's frame in the case of a static wormhole with the
comoving frame for the evolving geometry. Then one carries out a
simple Lorentz transformation to go into the frame of the
traveller.  The Riemann tensor components are obtained in this
traveller's frame and they lead to the tidal force constraints.
A similar analysis also holds for the acceleration constraint.
The major difference is that all constraints depend on time and
one has to find by extremization the time at which the
inequalities yield the most stringent condition. After this
extremization with respect to time one has to extremize w.r.t
$r$ to obtain the final condition which when satisfied would
make the wormhole traversable.

Since evolving wormholes are nonstatic, the study of quantum
field theory in these backgrounds may result in particle
creation. The relevant calculation would be fruitful to pursue.
It requires, however, the exact solutions of the scalar
wave/Maxwell or Dirac equations which at first sight may be
rather difficult to obtain. Numerical analyses can nevertheless
be done to provide hints into the number density and
distributions of the created particles.

\vspace{.3in}

{\bf ACKNOWLEDGEMENTS}

\vspace{.2in}

The authors are grateful to Matt Visser for allowing them to
quote one of his arguments and also for useful discussions.  SK
thanks Institute of Physics, Bhubaneswar for financial support
in terms of a fellowship.

\pagebreak

\vspace{1in}

{\widetext \centerline{\bf FIGURE CAPTIONS}}
\vspace{.6in}

{\widetext {\bf Fig. 1} : { F(t) vs t for $\Omega (t)=\sin {t} +
a \qquad (a)\omega =1, a=1.5 (b)\omega =1, a=2 (c)\omega =1, a=6
$}}

\vspace{.4in}

{\widetext {\bf Fig. 2} : {F(t) vs t for $\Omega (t) =\left
({{t^{2}+a^{2}}\over{t^{2}+b^ {2}}}\right )
\qquad b>a>0 (a)a=1, b=1.5 (b) a=1, b=2 (c)a=1, b=3  $}}

\end{document}